\documentclass[12pt]{article}
\usepackage{graphicx}
\usepackage{amssymb}
\usepackage{amsmath}
\usepackage{epsf}

\input{epsf}
\newlength{\extraspace}
\newlength{\extraspaces}
\newcommand{\figref}[1]{\protect\ref{#1}}

\makeatletter\@addtoreset{equation}{section}\makeatother

\newcommand{\ba}{\begin{eqnarray}
\addtolength{\abovedisplayskip}{\extraspaces}
\addtolength{\belowdisplayskip}{\extraspaces}
\addtolength{\abovedisplayshortskip}{\extraspace}
\addtolength{\belowdisplayshortskip}{\extraspace}}

\newcommand{\ea}{\end{eqnarray}}

\newcommand{\E}{{\cal E}}
\newcommand{\M}{{\cal M}}
\newcommand{\calO}{{\cal O}}

\newcommand{\N}{{\cal N}}
\newcommand{\F}{{\cal F}}

\newcommand{\txi}{{\tilde \xi}}

\newcommand{\IZ}{\mathbb{Z}}

\newcommand{\be}{\begin{equation}
\addtolength{\abovedisplayskip}{\extraspaces}
\addtolength{\belowdisplayskip}{\extraspaces}
\addtolength{\abovedisplayshortskip}{\extraspace}
\addtolength{\belowdisplayshortskip}{\extraspace}}
\newcommand{\ee}{\end{equation}}

\newcommand{\im}{{\rm Im \,}}

\def\del{\partial}
\def\Dslash{\not{\hbox{\kern-4pt $D$}}}
\def\dslash{\not{\hbox{\kern-2pt $\del$}}}

\def\de {\partial}

\begin{document}
\begin{titlepage}

\begin{center}

{\hbox to\hsize{ \hfill SLAC-PUB-11233}} {\hbox to\hsize{ \hfill SU-ITP-05/17}}

{\hbox to\hsize{ \hfill hep-th/0505208}}

\vspace{1.5cm}

{\Large \bf{A Stringy Test of Flux--Induced Isometry Gauging}}  

\vspace{1.5cm}

Amir--Kian Kashani--Poor$^{1,2}$ and Alessandro Tomasiello$^2$

\vspace{8mm}

$^1${\it SLAC, 2575 Sand Hill Road, Menlo Park, CA 94025} \\
$^2${\it ITP, Stanford University, Stanford CA 94305-4060}
\vspace*{1.5cm}

{\bf Abstract}\\

\end{center}

\noindent
Supergravity analysis suggests that the effect of fluxes in string theory compactifications is to gauge isometries
of the scalar manifold. However, isometries are generically broken by brane instanton
effects. Here we demonstrate how fluxes protect exactly those isometries 
from quantum corrections which are gauged according
to the classical supergravity analysis.  We also argue that all other 
isometries are generically broken.

\end{titlepage}

\section{Introduction}
Dimensionally reducing 10d supergravity on a Calabi-Yau manifold in a background of fluxes gives rise to gauged supergravity in four dimensions \cite{Polchinski}. The `gauged' 
refers to the fact that the hypermultiplets are charged under some of the vector multiplets. Since the hypers coordinatize a complicated scalar manifold, they can only acquire charges if the metric on this manifold exhibits isometries, which can then be gauged. This is analogous to gauging the shift symmetry of a free scalar field, which we can think of as gauging the translational isometry of the Euclidean line. The scalar manifold coordinatized by hypermultiplets in $\N=2$ supergravity is a quaternionic manifold \cite{baggerwitten}. When the supergravity arises upon CY compactifications of string theory (we will be considering IIA in this paper), the scalar fields in the hypermultiplets arise from complex structure moduli of the CY, as well as from the RR 3-form potential. Isometries of the scalar manifold arise (roughly) due to the shift symmetry of the RR potential in the 10d action. The 10d supergravity action in turn has this property because it captures the low energy interaction of strings, and these are not charged under the RR fields. The non-perturbative spectrum of string theory, on the other hand, does contain objects, D-branes, that are charged under these fields. 
We hence expect D-brane instantons  to break isometries of the quaternionic metric by introducing explicit RR potential dependence into the action. How must the fluxes--gauge-isometries prescription be modified to incorporate the lifting of these isometries? To give the punch line away, not at all. Our analysis shows that permissible instantons are correlated with fluxes in such a way to protect those isometries which are to be gauged from being removed by quantum corrections. 

This does not mean that the only way the flux modifies the action 
is by gauging. Indeed, in the second part of this paper 
we develop the case that other isometries of 
the hypermultiplet moduli space are lifted by instanton corrections which are themselves flux dependent. With the current state of brane instanton technology, it is a daunting task to show that these corrections preserve the quaternionic structure of the metric. The underlying assumption however is that fluxes break supersymmetry spontaneously in 4 dimensions, in which case this is guaranteed.

\section{Isometries protected by fluxes}
\label{sec:e2}
In this section we show how NS flux protects certain isometries. 
In the first two 
subsections we briefly remind the reader how the gauging arises upon 
dimensional reduction, and how brane instantons enter the calculation.
In subsection \ref{sec:lost}, we discuss how  
quantum corrections seem to invalidate the fluxes--gauge--isometries picture, 
by removing the isometries. Tadpole considerations come to the rescue 
constraining allowed instantons so that the isometries to be gauged are 
preserved. In this last subsection, we also introduce some instanton configurations which will occupy us in 
later
sections, when trying to assess how the other isometries are affected 
by brane instanton corrections.

\subsection{Review of fluxes gauge isometries}
We consider type IIA compactified on a Calabi-Yau $X$. 
Each hypermultiplet, aside from the universal one, contains two scalars $z^a$ which stem from complex structure moduli of $X$, and two 
scalars $\xi^A$, $\tilde \xi_A$ from the coefficients of $C_3$ 
in an expansion in a basis for $H^3$: $C_3 = \xi^A \alpha_A + \txi_A \beta^A$. 
Here $a=1, \ldots, h^{2,1}$, $A=0,\ldots, h^{2,1}$. $\xi^0$ and 
$\tilde \xi_0$ together with the dilaton $\phi$ and the NSNS axion $a$
are the scalars of the universal hypermultiplet. 
Via dimensional reduction, these scalars enter into the metric of the hypermultiplet 
scalar manifold as follows \cite{ferrara1,ferrara2}:
\ba
ds^2 &=& d \phi^2 + g_{a\bar b} dz^a d\bar z^{\bar b} + \frac{e^{4\phi}}{4}\left[da + \txi_A d\xi^A - \xi^A d\txi_A\right]\left[da + \txi_A d\xi^A - \xi^A d\txi_A\right]  \nonumber \\
& &-\frac{e^{2\phi}}{2} (\im \M^{-1})^{AB}\left[d\txi_A  +\M_{AC}d\xi^C \right] 
   \left[d\txi_B  +\overline{\M}_{BD}d\xi^D \right] \,. \nonumber
\ea
We consider turning on the fluxes $F_4= e_i \tilde{\omega}^i$ and $H = p^A \alpha_A + q_A \beta^A$, with $\tilde{\omega}^i$ a basis for $H^{2,2}(X)$ (this 
essentially encompasses the most general choice of fluxes consistent with locality of the
four--dimensional action \cite{lm,kk}).
Turning on $F_4$ leads to the gauging of the isometry
\ba
(k_F)_i = -2 e_i \partial_a  \label{isometryF}
\ea
of the scalar 
manifold by the $i^{th}$ vector multiplet, while $H$ leads to the gauging of
\ba
k_H &=& (p^A \txi_A - q_A \xi^A) \partial_a + p^A \partial_{\xi^A} + q_A \partial_{\txi_A}     \label{isometryH}
\ea
by the graviphoton.

In the following, the symplectic structure on $H^3$ will not be relevant for our considerations, hence we do not indicate it in our notation, writing $C_3 = \xi_i \gamma^i$, for $\{ \gamma^i \}$ a basis of $H^3$, and likewise $H = p_i \gamma^i$. We also introduce the dual basis in homology $\{ \Gamma_i \}$, such that $\int_{\Gamma_i} \gamma^j = \delta_i^j$.

\subsection{Brane instantons} \label{braneinst}
Brane instantons are Euclidean branes which wrap cycles in the internal manifold at a point in time. The rules of incorporating brane instanton corrections to 4d effective actions are not completely understood. Arguing along the lines of \cite{bbs,hm,moore}, we use the E-brane action to derive vertex operators which describe the coupling between spacetime fields and brane fields. To calculate instanton contributions to a spacetime correlator, we assemble the appropriate vertex operators and then perform the path integral over the brane degrees of freedom. In addition, any contribution in the instanton sector will contain a factor from the classical part of the brane instanton action, 
\ba
\nonumber \langle \calO \rangle_{inst} &=& e^{-S_{inst, class}} \ldots \\
&=& e^{-vol + i \int C} \ldots \,.
\ea

\subsection{Isometries Lost and Isometries Regained}
\label{sec:lost}
The isometries $k_F$ and $k_H$ we are considering depend on modes descending from the NSNS $B$ field and the RR potential $C_3$.
$da$ derives from the spacetime components of $B$. Since these components of $B$, in the absence of sources and non-trivial spacetime topology, are pure gauge
and can hence be set to zero in the brane action, neither a contribution to $S_{inst, class}$ depending on $a$ nor a vertex operator coupling $a$ to degrees of freedom on the brane arise.\footnote{The underlying assumption here is that going off-shell does not give rise to additional vertex operators.} 
 We hence do not worry about brane instanton corrections lifting $k_F$, and in the case of $k_H$, we can focus on the shift symmetry part of the isometry in $\xi_i$.\footnote{Because the isometry in question involves a field dependent shift in $a$ in addition to a constant shift in the $\xi_i$, the shift symmetry of RR potentials does not prove that it is preserved by {\it perturbative} stringy corrections. Nevertheless, this is believed to be the case \cite{strominger, antoniadis1, antoniadis2}. We feel the question of non-perturbative lifting is more pressing, as brane instantons pose an immediate threat to the isometry's survival.}

For E2 branes (the instantonic
version of D2 branes), the classical contribution already appears to 
generically break any isometry involving a shift symmetry in the 
fields $\xi_i$.
Consider an instanton configuration consisting of E2 branes  wrapping various 
cycles; let the sum of these cycles in 
homology be 
\ba \label{e2config}
\Gamma_{inst} &=& \sum_i c^i \Gamma_i  \,.
\ea
This configuration contributes a $\xi_i$ dependence
\ba \label{cx}
\int_\Gamma C_3 = \sum_i c^i \xi_i
\ea
to the effective action. Acting with (\ref{isometryH}) on (\ref{cx}) 
results in 
\begin{equation}
  \label{eq:break}
k_H (\int_\Gamma C_3) = \sum_i c^i p_i  \ ,
\end{equation}
which generally does not vanish. 
Hence the fluxes--gauge--isometries picture does not appear to survive quantum
corrections, since the isometries to be gauged do not. We now proceed to 
show why this is not the case. 

The key point is that $H$ flux induces a magnetic charge for the gauge field 
on the brane. One way to see this is to note that $B$ always appears in the 
combination $F+B$. The Bianchi condition thus becomes $dF=0 \rightarrow 
dF+H=0$ (in general, this has a refinement in integral cohomology which 
gains a contribution from the third Stiefel--Whitney class; however, for 
three--cycles, this is always zero; see for example \cite{bbs}). 
Hence, an E2 brane wrapping a compact 
cycle threaded by $H$ flux 
violates Gauss' law on its worldvolume, unless $H$ is cohomologically trivial. 
On an instanton wrapping 
a cycle $\sum c^i \Gamma_i$, in the notation introduced above,
this condition reads $\sum_i c^i p_i=0$. The LHS of this equation is exactly the potential violation of 
the isometry which we obtained in (\ref{eq:break})! We conclude that 
the instantons which would break the isometry $k_H$, i.e. those for which $\sum_i c^i p_i \neq 0$, are ruled out by Gauss' law.

\begin{figure}[h]
\begin{center}
\resizebox{5cm}{!}{\includegraphics{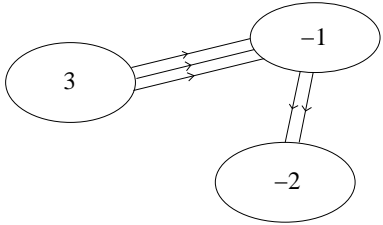}}
\end{center}
\caption{\small{A permissible instanton configuration. The ellipses denote E2 
branes wrapping homologically different 3 cycles, the connecting lines are E0 
branes. The numbers inside the ellipses indicate the quanta of  $H$ flux 
through the respective cycle.}}\label{e2d0}
\end{figure}

One could think of more elaborate strategies to produce instantons
which break the isometry.  
We can add E0 branes ending on the E2 brane, as these are magnetic sources 
for the worldvolume gauge field $F$. A configuration of E2 and E0 branes which 
is consistent with Maxwell's law is depicted in figure \figref{e2d0} 
(this configuration is a close relative of ones which have appeared in 
\cite{mms}).
However, each E0 is a charge on the E2 on which it starts, but a charge of 
opposite sign on the E2 on which it terminates. Thus we can quickly 
conclude that also for this more general configuration 
\begin{equation}
  \label{eq:cp}
  \sum_i c^i p_i=0\ .
\end{equation}

After having saved the isometries we needed from quantum corrections, we
will now show how non--academic the worry was: in general, all the
isometries which are not needed for gauging get lifted by quantum corrections.

\section{Isometries lifted by instantons}
In the previous section, we demonstrated that those isometries of the form
\ba
k_H &=& (p^A \txi_A - q_A \xi^A) \partial_a + p^A \partial_{\xi^A} + q_A \partial_{\txi_A}  \,.   \label{isometry}
\ea
that are to be gauged upon turning on $H$ fluxes are protected. More specifically, we argued that the part of the isometries at stake is the shift symmetry of the instanton contribution in $\xi_i$ ($i=1,\ldots, b_3$, $\xi_i$ collectively denoting $\xi^A$ and $\txi_A$). These shift symmetries 
$\{\partial_{\xi_i} \}$ span an integral 
lattice ${\cal S}_{b_3}$ of dimension $b_3$ (the coefficients are 
integer). The $H$-flux represents a vector in this lattice, and we showed 
that symmetries in the direction of this vector are protected by Gauss' law. In this section, we wish to show that the symmetries in the $b_3-1$ dimensional orthogonal complement ${\cal S}_{b_3-1}$ are lifted.
 
\subsection{Which instantons are present?}
In the previous section, we considered instantons only at the crude level of homology. It is far from obvious that choices of representatives of the homology cycles appearing in the instanton configurations of the previous section can be made that give rise to a local minimum of the action. Since we were able to rule out the lifting of the isometries of interest by all such candidate instantons, the possible over-counting was inconsequential. In this section, we wish to argue that all isometries that are not protected by flux are lifted. We must hence consider the question of what type of brane configurations 
qualify as instantons more carefully. 

In the absence of flux, if we can saturate the supersymmetric BPS 
bound, we can minimize the action globally;
geometrically, this gets translated into the condition that there exist
a special Lagrangian (SLag) submanifold representative of the cycle with a flat
bundle over it, as first obtained in \cite{bbs}. The presence of a $B$ field modifies this
result in two ways \cite{Marino:1999af}. First, in the absence of E0 branes, 
$H$ has to be cohomologically trivial on the 
submanifold, as we have already seen. Second, the flatness condition now
becomes, not surprisingly, $F+B=0$. In presence of $B$ 
the bundle on a brane should not be seen as a true bundle anymore, but as
a projective bundle determined by $B$ (see for example \cite{kapustin}); 
hence this condition can be satisfied. We have pointed out before that 
the integral refinements to the tadpole condition are not relevant for 
three--dimensional SLag's. 

To lift all elements of ${\cal S}_{b_3-1}$ by E2 instantons wrapping SLags would require a basis of the subspace of 3-homology dual to ${\cal S}_{b_3-1}$ for which each element had a SLag representative {\it somewhere} in moduli space. The final qualification is possible because the existence of isometries is a global question on moduli space; it is sufficient
that the vector field not be Killing at {\it some} point to declare the 
isometry broken. In spite of this qualification, this is a rather strong requirement, since by appropriate choice of the flux $H$, the corresponding ${\cal S}_{b_3-1}$ can be dual to any codimension 1 subspace of the third homology vector space $H_3$; and it is not clear that {\it all} cycles in $H_3$ 
(not just a basis\footnote{That such a basis exists can be seen by going 
to the mirror. There, we have to look for a basis of K--theory all elements of
which are $\Pi$--stable \cite{dfr}. Thanks to the Lefschetz theorem, for any
integral element in $H^{1,1}$ and in $H^{2,2}$ there exists a dual analytic 
cycle. These cycles are all stable in the large volume limit of the mirror. 
Hence their duals are SLag at least at one point of the complex structure 
moduli space of the original Calabi--Yau.}) have a SLag 
representative. 
We therefore argue next for the existence of a class of instantons, of the type depicted in figure \ref{e2d0}, which, to lift all elements of ${\cal S}_{b_3-1}$ for any choice of flux $H$, requires only that a basis of the whole of $H_3$ exists such that each element has a SLag representative somewhere in moduli space.

To argue for the existence of such instantons, let us first consider the fluxless case. We start by a brane wrapping a SLag cycle at a certain point in moduli space and follow its fate as we move in moduli space \cite{km} . There are stability
walls beyond which a SLag which represents a cycle $C=A+B$ 
can decay to two separate but intersecting 
SLag's, one on the cycle $A$ and one on the cycle
$B$, which are separately BPS, but not mutually BPS.
Conversely, on the other side of the wall, $A$ and $B$ put together
decay to the SLag representing $A+B$. We can think of the process as a formation 
of a bound state, via condensation of a (string) world--sheet tachyon.

Now, in our case in which $H$ is also present, after the SLag representing
$A+B$ decays, we are left with two intersecting SLag's representing $A$ and $B$, such that,
if we previously had $\int_{A+B} H=0$, now $\int_{A,B} H$ need not be zero. E0's must then be added to the configuration to rescue Gauss' law.
We have thus almost arrived in the situation in figure \ref{e2d0}; except that the ellipses
here intersect. The E0's can all shrink to the intersection point, or be stuck
at some geodesics between the cycles which is a local minimum of the 
path--minimization problem. That these configurations contribute is
required by the fact that amplitudes to which instantons contribute cannot abruptly jump on any 
stability wall, so when a SLag wrapped by an E2 brane disappears, something else must take its place. 
Encouraged by this, we can now ask whether action minimizing E2--E0 configurations 
also arise for E2's wrapping individual SLag cycles $A_i$ that do not intersect, hence cycles the homology class of the sum $C=\sum A_i$ of which does not 
necessarily have a SLag representative anywhere on moduli space. Again, in the intersecting case, connecting the configuration continuously to a BPS configuration was circumstantial evidence for this being the case.
In the non--intersecting case we have
to work harder. Each of the E2's we are taking is wrapping a SLag, and each E0, now of finite length, lies
on a geodesic connecting them, so each constituent separately minimizes the 
action; but can there be a deformation of the whole system that lowers the 
action? We do not have a definite argument to rule this out, but we can note
the following. Given a typical size $R$ of the Calabi--Yau, the E2's have
action of order $R^3/(g_s l_s^3)$, whereas the E0's have action 
$\sim R/(g_sl_s)$.
To speak of geometric objects in the first place, we have to take $R\gg l_s$; 
hence the E2's contribute much more action than the E0's do. Hence it makes
sense to first settle the E2's in such a way as to minimize the action, and
then insert the E0's to accommodate them. This intuition will also guide
us in the zero mode analysis which we perform below.

Of course we do not have full control over this new kind of instanton; we will 
try to analyze the contribution of such instantons together with that of the more conventional instantons in the rest of this section. We 
will argue that they do contribute; at the very least, it seems that the 
conditions for them to contribute have little to do with the conditions 
that a cycle admit a SLag representative.

\subsection{Ingredients of the instanton computation}
\label{sec:ingredients}

As in \cite{bbs}, we shall take contributions to the 4 hyperino coupling 
\cite{ferrara} $R_{IJKL} \bar{\chi}^I_+ \chi^J_+ \bar{\chi}^K_- \chi^L_-$ in the instanton background as our benchmark of whether instanton configurations are deforming the hypermultiplet metric ($R_{IJKL}$ is the $Sp(2h^{2,1})$ piece of the quaternionic curvature, $_\pm$ indicate the chirality of the spinors). For this purpose, we consider the correlator $\langle \bar{\chi}^I_+ \chi^J_+ \bar{\chi}^K_- \chi^L_- \rangle$ in an instanton background. As we noted in section (\ref{braneinst}), {\it any} contribution to this correlator from a given instanton sector will involve a factor of the classical instanton action, $S_{inst, class} = vol + i \sum c_i \xi^i$, and hence break isometries of the quaternionic metric that are not preserved by $\sum c_i \xi^i$.

As we mentioned in that section, the rules for performing brane instanton calculations are not well understood. Results for the exact form of instanton corrections to the hypermultiplet metric can be found in the literature \cite{vafa, Becker, Gutperle, Theis, Ketov, Davidse, rocek, stefan}; they are usually obtained via indirect arguments, or in 4d SUGRA. Our modest goal is
to determine whether the contribution to the correlator in an instanton 
background vanishes or not; hence, any non-vanishing contribution will do. 
When arguing for the presence of corrections to correlators in 
non--trivial instanton backgrounds, special attention must be devoted to two issues: fermionic zero modes,
and integration of non--trivial determinants. 
The issue of zero modes will be the topic of the following subsection. The second issue we consider here. 

After having integrated out massive modes, one gets an effective action
to be integrated on the manifold of bosonic zero modes, and which is 
a function of them only. In many cases with appropriately high
supersymmetry, this function can simply be 1, but in our case supersymmetry is spontaneously
broken (a fact which will play an important role later). So one 
should worry that integration of this effective action gives zero. This worry does not arise for instantons that are isolated, such that the integral over the bosonic moduli space is simply a spacetime integral. For this reason we will 
restrict to the case in which there exists a basis
$\{\Gamma_i\}$ of $H_3$ consisting of homology spheres, which are isolated. This turns out
to be the case whenever we have a Landau--Ginzburg point in the moduli space,
though not in general.
Here we are also implicitly using that zero modes of the combined system E2--E0
can be obtained by considering the E2's alone, a fact for which we will
argue in section \ref{sec:zeropart}.

\subsubsection{Zero modes}

In this section, we discuss the notion of exact zero modes, to argue that in contradistinction to the fluxless case, any brane instanton configuration can contribute to corrections to the hypermultiplet metric. In a nutshell, this is because the fluxes break the supersymmetry of the background. In the latter part of this section, we introduce the notion of approximate zero modes as a useful tool for organizing our perturbative calculation.

A fermionic path integral is formally defined by expanding the fermion $\psi$ in modes $\psi_i$, $\psi = \sum \xi_i \psi_i$, and performing Berezin integration over the Grassmann coefficients $\xi_i$. The $\psi_i$ are, as we will discuss, the eigenmodes of a suitably chosen operator $M$.
If the integrand does not depend on one of the $\xi_i$, the integral vanishes. Assume the action is quadratic in $\psi$, $I = \int \psi D(\phi) \psi$, with $\phi$ generically denoting bosonic fields in the theory. If the classical equations of motions have a solution $(\phi_0, \psi_0)$, such that in particular $\psi_0$ is a non-trivial element of the kernel of the operator $D(\phi_0)$, then we call $\psi_0$ an exact fermionic zero mode. Such exact zero modes will generically not exist. In supersymmetric theories, they can be generated by acting on bosonic instanton solutions with the supercharges which are broken in the presence of instantons. Expanding $\psi$ in modes of $D(\phi_0)$, i.e. choosing $M=D(\phi_0)$, we see that the action evaluated at $\phi = \phi_0$ is independent of the exact zero modes. The restriction $\phi = \phi_0$ implies that we can conclude that without insertions to soak up the exact zero modes, the path integral vanishes {\it at tree level in all bosonic fields}. Zero mode dependence does arise away from $\phi = \phi_0$, i.e. when quantum modes of the bosonic fields are taken into account. To decide whether these quantum modes are sufficient to lift the zero modes is tricky. ADS argue that aside from those zero modes generated by supersymmetry, this should be possible.\footnote{A more modern localization argument for the supersymmetric zero modes not being lifted relies, roughly speaking, on introducing a different mode expansion of the fermion over every point of the bosonic field space.  This procedure can be realized in the presence of a fermionic symmetry $\F$ acting on field space $\E$. If the action is free, i.e. is without fixed points, bosonic field space is given by $\E/\F$, and the fermionic modes over each point of $\E / \F$ are obtained by acting by $\F$. As long as the integrand is invariant under the action of $\F$, it will be independent of these modes, and the fermionic integral over each point of $\E / \F$ vanishes.
 If however there are
loci on which the action is not zero, i.e. supersymmetric field configurations, then these must be excluded from $\E$ before the quotient can be taken. These loci give rise to non-trivial contributions to the path integral.
BPS instantons are field configurations which are fixed points of some but not all of the supercharges. Such configurations can contribute to the expectation value of operators which are not invariant under those supercharges which do not fix the instanton. 

For example, if we want to determine whether a superpotential is generated in a four dimensional theory, we can calculate the correlation function of two fermions. The two fermionic insertions will break two of the supercharges. The 
other two supercharges localize the path integral on the zero locus of the two associated fermionic symmetries. Since the BPS instanton is contained in this zero locus, the generation of a superpotential due to this instanton is possible. A similar reasoning can be applied to $\int d\theta^4$ terms in 
${\cal N}=2$ theories.}
Even without performing a technically daunting detailed analysis for our instantons, we hence feel doubly assured that they will generically contribute to the hypermultiplet metric: exact fermionic zero modes will generically not exist, since supersymmetry is broken by the fluxes; if they do exist, they will be lifted by quantum interactions.

Since we have now convinced ourselves that our amplitude does not vanish due 
to known non-perturbative arguments, we can go about organizing our 
perturbation theory. In the large volume limit, form fields are suppressed
by powers of the size of the Calabi--Yau $R$. 
 The flux of a $p$--form field $F_p$, 
which is an integral over a cycle, is of order $F_p R^p$, and has to be 
an integer. So $H$ is of order $1/R^3$ and $F_4$ of order $1/R^4$.
In the limit in which $R$ is large, fluxes can hence be treated as 
perturbations. We decompose the quadratic operator $D$ introduced in 
the previous section as $D_0 + \delta D_0$, where all of the flux dependence is incorporated into $\delta D_0$, which we treat as an interaction term. Zero modes of $D_0$ (which are present due to the supersymmetry in the absence of fluxes) are called approximate zero modes. By expanding fermions in modes of $D_0$, i.e. choosing $M=D_0$, we need to soak up these approximate zero modes in each 
term in our perturbative expansion, by insertions or by pulling down $\psi \delta D_0 \psi$ interactions. We flesh out this picture in the remaining parts of this section.

\subsubsection{Instanton vertices}
\label{sec:zeropart}
\paragraph{E2 instantons}
For E2 instantons wrapping SLags, the fermion that featured prominently in the above is the world volume fermion $\theta$ on the E2. It has $4$ approximate zero modes, as in the absence of flux, the instanton breaks half of the 8 supersymmetries of the background. Any correlator of spacetime fields hence needs $4$ insertions of $\theta$ so as not to vanish in this instanton background. These insertions are provided by bringing down interaction vertices of the worldvolume action involving $\theta$ (these will be gravitino-$\theta$ couplings and flux induced mass terms for $\theta$, as we discuss below). Let us label these interaction vertices by ${\cal V}(x)_i$. $x$ here denotes the position of the instanton in spacetime. It is a collective coordinate the integration over which is included in the path integral measure. To soak up the correct number of zero modes, we hence end up with a factor $\int dx \, \prod_{\alpha} {\cal V}_{i_\alpha}(x)$ in the path integral. As far as Feynman rules are concerned, such a factor behaves just like an ordinary interaction vertex in the action. 

\paragraph{E2-E0 instantons}
We wish to argue that the above considerations are also sufficient to deal with the E2--E0 instantons. For this purpose, we want to consider, in addition to the flux, the
E0's as a perturbation as well. This seems consistent, since, 
as we have seen above (section \ref{sec:ingredients}), the E0's are much 
lighter than the E2's. We hence want to think of the E0 branes as solitonic solutions in the E2 brane world volume theory. This has been worked out in various situations dual to the E2--E0 system of interest here in the literature (see e.~g.~\cite{diaconescu, agit, Callan:1997kz}).
The E0 branes introduce magnetic charges for the gauge field on the world volume. Solutions to the classical equations of motion for the bosonic fields $X_i$ which encode the transverse position of the brane in the presence of these charges will, far away from the E0 sources, simply be $X_i =\mathrm{const}$, while close to the sources, $X_i$ will spike out to build a bridge to the other E2 brane. In this picture, the E0 brane degrees of freedom arise as fluctuations around the bridge part of the configuration of the world volume fields of the E2. These could be studied by replacing this contribution to the E2 brane action by a one dimensional E0 brane action.

With this picture in place, we want to perform a zero mode analysis of the E2--E0 system. First consider bosonic fluctuations. These should be well approximated by independent fluctuations of the E2 branes, with the E0's simply adjusting themselves to accommodate these fluctuations: the E0's, being geodesics, will generically be rigid far away from the junction points with the E2's. The fluctuations hence have support mainly on the individual E2's. The same is hence true for the approximate fermionic zero modes related to these via supersymmetry. We will assume that these are the only zero modes present. This point of view instructs us to consider action minimizing configuration where each E2 separately, i.e. ignoring the E0 branes and by consistency therefore also the potentially cohomologically non-trivial $H$-flux, minimizes the action, e.g. by wrapping a SLag, which is the case we will consider. Then, while they generically will break (in the absence of flux) different supersymmetries, each should exhibit a world volume fermion $\theta$ with four approximate zero modes (arising from the action of different supercharges on the E2's). Again, these
zero modes are the ones that would be exact if we were considering the E2's 
only in a background without flux, and become approximate once we add the 
perturbation of the flux and the E0's.

Aside from allowing us to settle the E2 constituents on SLags, the above analysis suggests that the approximate zero modes of the E2--E0 system can be thought of as being supported on the E2's only.
We will hence only need to determine interaction terms of the E2 world volume fermion $\theta$.

\paragraph{Couplings of the hyperinos to brane fermions}

Following \cite{bbs}, we 
 obtain the coupling between the worldvolume spinor and the hyperinos
from the corresponding interaction term in the brane action.
For simplicity, we derive this coupling first in the context of 
reduction from eleven dimensions to five, that is, by considering the 
worldvolume action of the M2 brane. We will then reduce the result from
five to four dimensions.

In the CY reduction of 11d SUGRA to 5d $\N=2$ SUGRA, the hyperinos descend 
from the 11d gravitino.
In the M2 action there is a coupling \cite{Bergshoeff,bbs,norcor,marolf} between the gravitino
$\psi_M$ and the worldsheet spinor $\theta$:
\begin{equation}
  \label{eq:m2}
  \int \psi_M^\dagger \left( \de^a X^M \de_a X^N \gamma_N - \frac i2 
\epsilon^{abc} \de_a X^M \de_b X^N \de_c X^P \gamma_{NP} \right) \theta=
\int \psi_a^\dagger \gamma^a (1-\Gamma)\theta\ .
\end{equation}
The first term 
stems from expanding out the superspace counterpart of $\sqrt{g}$, the second 
term from the WZ term for $C_3$ on the M2 brane. We have also rewritten 
the vertex in a more compact form, anticipating some notation that will be
useful later; quantities with worldsheet index $a$ are pulled back using
the $\partial_a X^M$, as usual, for example $\psi_a\equiv 
\partial_a X^M \psi_M$. The matrix $\Gamma\equiv \frac i{3!} 
\epsilon^{abc}\gamma_{abc}$ plays a prominent role in kappa--symmetric actions,
and it will appear again in flux--induced vertex operators below. 

To extract the hyperino-$\theta$ coupling from the gravitino-$\theta$ vertex, we need to consider the mode expansion of the gravitino.
In compactifying M-theory on a CY, the massless hyperinos 
$\chi_I$, $I=2,\ldots, 2h^{2,1}+2$, arise as coefficients of the expansion 
of the internal directions of the 11d gravitino  (as well as the gauginos $\eta^\Theta$,
$\Theta=1, \ldots, h^{1,1}-1$; recall that, in contrast 
to Calabi--Yau reduction to four
dimensions, one obtains $h^{1,1}-1$ vector multiplets; 
an additional vector multiplet 
arises upon reduction of the gravity multiplet from five to four dimensions).
The harmonic spinors in the internal 
directions are well--known to be given by $\oplus_p H^{0,p}(CY)$. A harmonic vector--spinor 
whose vector index is holomorphic is given by 
$\oplus_p H^{0,p}(CY,T)=\oplus_p H^{2,p}(CY)$; 
we have used the Dolbeault theorem and triviality of the canonical bundle. 
In formulas,
\ba
  \label{eq:dol}
 \psi_i&=& \sum_{\Lambda=1}^{h^{2,1}}\chi_\Lambda\otimes
d^\Lambda_{i\bar j \bar k} \gamma^{\bar j \bar k} \epsilon_+
+ \sum_{\Theta{}=1}^{h^{1,1}}\eta_\Theta{}\otimes
d^\Theta{}_{i\bar j} \gamma^{\bar j}\epsilon_+\\
\nonumber &=&\sum_{\Lambda{}=1}^{h^{2,1}}\chi_\Lambda\otimes
d^\Lambda_{i\bar j \bar k} \gamma^{\bar j \bar k} \epsilon_+
+\chi_0\otimes
d^{h^{1,1}}_{i\bar l} \Omega^{\bar l}{}_{jk}\gamma^{j k} \epsilon_-
+\sum_{\Theta{}=1}^{h^{1,1}-1}\eta_\Theta\otimes
d^\Theta_{i\bar j} \gamma^{i\bar j}\epsilon_+\,,
\ea
where we have chosen the basis of $H^2$ such that 
$d^{h^{1,1}}\sim J$ and set $\chi_0=\eta_{h^{1,1}}$.  
Here $\epsilon_\pm$ are the two Weyl
covariantly constant spinors on the Calabi--Yau.
We can likewise reduce $\psi_{\bar i}$,\footnote{In Minkowski signature, 
the gravitino is Majorana, which leads to a relation between $\psi_i$ and
$\psi_{\bar i}$. However, in Euclidean signature, which is 
appropriate to instanton computations, the Majorana condition is
not possible in 11d; so we will always work with a complexification of the 
fermionic degrees of freedom. This subtlety is present in many 
instanton computations, also in four dimensions. For an analysis of rigid $\N=2$ SUSY in Euclidean signature, see \cite{Cortes}.}
 getting another set of 5d fermions
$\{ \tilde\chi_\Lambda\}$. 
Then, by defining 
$d^0_{ijk}\equiv d^{h^{1,1}}_{i\bar l} \Omega^{\bar l}{}_{jk}$ and 
$d^0_{\bar i \bar j \bar k}\equiv d^{h^{1,1}}_{\bar i l} 
\Omega^{l}{}_{\bar j\bar k}$
we can assemble everything in the form \cite{bbs}, 
\ba
\label{modes}
\Psi_M &=& \chi_I \otimes 
d_{MNP}^I \gamma^{NP}(\epsilon_+ + \epsilon_-) + \ldots \,,
\ea
$I=1, \ldots, 2h^{2,1}+2$, where $\chi_I= \{ \chi_\Lambda, \tilde \chi_\Lambda \}$, $d^I = \{ d_{3,0}, d^{\Lambda}_{2,1}, d^{\Lambda}_{1,2}, d_{0,3} \}$,  
and $\ldots$ denotes the gaugino part. The zero modes of the world volume spinor $\theta$ are of the form $\xi \otimes \epsilon$, where $\xi$ is a 4d spinor and $\epsilon = e^{i \phi_{\Gamma}}\epsilon_+ - e^{-i \phi_{\Gamma}}\epsilon_-$  \cite{bbs}; $\phi_\Gamma$ here denotes the phase of the SLag.
Inserting the expansion (\ref{modes}) into the gravitino-$\theta$ vertex yields the following hyperino--$\theta$ vertex operator \cite{bbs}.
\ba
{\cal V}_{\chi \theta} = 8i \chi^{I\dagger} \xi \int_{\Gamma} 
e^{\pm i\phi_\Gamma}d_I \,,
\ea
with the positive sign for $I=1,\ldots, h^{2,1}+1$, and the negative sign for the remaining indices. For notational convenience, we redefine the basis of three cocycles to absorb the $\phi_\Gamma$ phase.
When we reduce this expression to four dimensions, it retains the same
form, $\chi_I= \chi_I^+ + \chi_I^-$ being understood now as a 
four--dimensional Dirac spinor. 

\paragraph{The flux induced mass term}
The quadratic term in the world volume fermion in the M2 action is given by 
\cite{norcor, marolf, renatasorokin} as
\ba
\int\theta^\dagger 
\left[
\gamma^a D_a + 
\frac1{288}G_{MNPQ}(\gamma_a{}^{MNPQ} +8\delta_a{}^M \gamma^{NPQ})\gamma^a
\right](1-\Gamma)\,\theta{}\ .
\ea
As above, $a,b \ldots$ are coordinates along the world volume of the brane, 
$M,N \ldots$ are target space indices, and $\Gamma=\frac i{3!}\epsilon^{abc}
\gamma_{abc}$. The gamma matrix combination in the 
bracket is familiar from the gravitino supersymmetry variation of 
11--dimensional supergravity. 

We now perform the computation of the instanton vertex induced by flux in 
the following way. While staying in eleven dimensions, we take $G=
F + H\wedge dx^5$; this gives rise to a vertex for five--dimensional
spinors which we then reduce along the fifth direction 
to four--dimensional Dirac spinors. Since $\gamma^{11} = \gamma_5^{5d} \otimes \gamma_7^{6d}$ arises in the part of calculation involving $H$, the result depends on the chirality of the 4d part of the world volume spinor zero modes. The final result reads
\ba
\label{eq:fluxvertex}
{\cal V}^\pm_{\theta \theta} =
\xi_\pm^\dagger \xi_\pm\int\left[ -  \frac{1}{2} \mathrm{vol}_3 F\llcorner(J^2) \mp 
\frac13 \mathrm{Im}\Big(2H^{3,0}+ 6 H^{2,1} \Big)\right]
\,.
\ea

\subsection{Contributions to the curvature--4--fermion coupling}
We now assemble the pieces from the above analysis to calculate the leading instanton correction to the 4 chiralino term 
\ba
\langle \bar{\chi}^I_+ \chi^J_+ \bar{\chi}^K_- \chi^L_- \rangle  \label{4chi}
\ea
in the 4d action. 

In the case of an E2 brane wrapping a SLag, we have the same number of approximate zero modes as \cite{bbs}, namely 4. The leading instanton contribution to (\ref{4chi}) is hence given by four insertions of ${\cal V}_{\chi \theta}$, yielding
\ba
\langle \bar{\chi}^I_+ \chi^J_+ \bar{\chi}^K_- \chi^L_- \rangle_{inst} &\sim& 
e^{vol(\Gamma) + i\int_\Gamma C_3} \int_\Gamma d_I
\int_\Gamma d_J \int_\Gamma d_K\int_\Gamma d_L
\ea
as in \cite{bbs}. 

We will have additional zero modes for the E2--E0 configurations. Let us for definiteness
consider a case in which there are two SLag's $\Sigma_1$, $\Sigma_2$ connected
by some E0's, and that there are only the eight zero modes coming
from broken supercharges. In addition to the vertex operators from the previous case, we now need two ${\cal V}_{\theta \theta}$ vertices to soak up the approximate zero modes of $\theta$. The result is
\begin{equation}
  \label{eq:finalcomp}
\langle \bar{\chi}^I_+ \chi^J_+ \bar{\chi}^K_- \chi^L_- \rangle_{inst} \sim
 e^{vol(\Gamma_1 + \Gamma_2) + i\int_{\Gamma_1 + \Gamma_2} C_3} \sum_{i,j \in \IZ_2}
{\cal V}_{\theta\theta}^+(\Gamma_i){\cal V}_{\theta\theta}^-(\Gamma_j)
\int_{\Gamma_{i+1}} d_I \int_{\Gamma_{i+1}} d_J \int_{\Gamma_{j+1}} d_K \int_{\Gamma_{j+1}}d_L  \end{equation}
where ${\cal V}_{\theta\theta}^\pm$ are given by (\ref{eq:fluxvertex}).

The conclusion of this section is that E2--E0 instantons give contributions 
which generically break all isometries not protected by flux (due to the 
$C_3$ dependence coming from the exponent). Also note that the ${\cal V_{\theta \theta}}$  vertices introduce 
explicit flux dependence in the quantum corrected quaternionic metric.

\medskip

{\bf Acknowledgements}
We would like to thank L.~Anguelova, P.~Aspinwall, A. Ceresole, G. Dall'Agata, S.~Kachru, Juan Maldacena,
J.~McGreevy, G.~Moore, M.~Ro\v cek, A.~Strominger, S.~Vandoren for
useful discussions and correspondence. This work was supported by NSF
grant 0244728. The work of AK was also supported by the 
U.~S.~Department of Energy under contract number DE-AC02-76SF00515.

 \end{document}